# **Flying doughnut terahertz pulses generated from semiconductor currents**

Kamalesh Jana[1*], Yonghao Mi[1], Søren H. Møller[1], Dong Hyuk Ko[1], Shima Gholam-Mirzaei[1], Daryoush Abdollahpour[1,2], Shawn Sederberg[3], Paul B. Corkum[1*]

[1]*Joint Attosecond Science Laboratory, University of Ottawa and National Research Council Canada, 25 Templeton Street, Ottawa, ON K1N 6N5 Canada*

[2]*Department of Physics, Institute for Advanced Studies in Basic Sciences (IASBS), Zanjan 45137-66731, Iran*

[3]*School of Engineering Science, Simon Fraser University, 8888 University Drive, Burnaby, BC, V5A 1S6, Canada*

[*]e-mail: *kjana@uottawa.ca (K.J.) ; pcorkum@uottawa.ca (P.B.C.)*

**The ability to manipulate the space-time structure of light waves diversifies light-matter interaction and light-driven applications. Conventionally, metasurfaces are employed to locally control the amplitude and phase of light fields by the material response and structure of small meta-atoms. However, the fixed spatial structures of metasurfaces offer limited opportunities. Quantum control of dynamic semiconductor currents can also launch a space-time coupled terahertz (THz) pulse, including a 'flying doughnut'. In contrast to conventional approaches to generating structured light beams, quantum control enables the amplitude, sign and even configuration of the generated fields to be manipulated in an all-optical manner. Near the center of the beam flying doughnut pulses exhibit longitudinal magnetic or electric fields that are spatially isolated from their partner electric or magnetic fields. Harnessing the isolated magnetic fields from such pulses could provide a powerful new tool for studying magnetic-field-induced dynamics in matter. Here, we apply bichromatic azimuthally-polarized vector light pulses to gallium arsenide to generate transient ring currents. By measuring the space-time structure of the radiated THz electric field, we demonstrate the generation of 'flying doughnut' terahertz pulses. Applying Maxwell's equations to the measured data enables calculation of the space-time structure of the isolated magnetic fields. As a first application, we detect absorption features from ambient water vapor on the spatiotemporal structure of the measured electric fields and the calculated magnetic fields. Quantum control is a powerful and flexible route to generating any structured light pulse in the THz range, while pulse compression of cylindrical vector beams is available for very high-power magnetic-pulse generation from the mid-infrared to near UV spectral region. Pulses such as these will serve as unique probes for spectroscopy, imaging, telecommunications and magnetic materials.**

Locally modifying the spatiotemporal character of light leads to intricate light beams that are often called "vector beams" or "structured light"[1,2]. With structured light, we open the potential for harnessing the topological properties of light and topology has applications in many research fields including optical tweezers, communications, particle control, ultrafast metrology, and super-resolution microscopy[1-6]. Recent advances in optical technology have enabled fine control over the spatial and temporal structure of light, independently. Structuring multiple properties of light simultaneously leads to unique classes of light and new regimes of light-matter interaction. The technology required to generate such beams is virtually non-existent. In 1996, Hellwarth and Nouchi[7] mathematically described a new light mode that is a single-cycle light pulse having toroidal topology, called 'Flying Doughnut' pulses, that are non-separable in space and time and contain strong magnetic or electric field components along their propagation axis[7]. Recently, there has been substantial interest in creating such space-time coupled toroidal light modes[8,9] in the THz region, where single-cycle pulses have been available for decades[10].

Despite the impressive progress in ultrafast optics and beam shaping technology, generation of such spatiotemporal light modes remains challenging mainly due to their complex beam structure, and our inability to sculpt the entire bandwidth of the pulses. A few approaches based on nanostructured metasurfaces have been used to generate spatiotemporally tailored THz pulses[8,11,12]. However, active manipulation of the THz modes using conventional metasurfaces is not possible as most metasurfaces are static with fixed functionalities once fabricated. Also, the metasurfaces cannot withstand high pump intensity beyond their thermal damage threshold. Hence, there are limited opportunities for achieving substantially higher THz conversion efficiencies using metasurfaces.

Quantum control is an effective way to generate and control currents in semiconductors by exploiting quantum interference between pathways connecting the same initial and final states[13-15]. In our previous experiments, we have demonstrated structured current generation by applying spatially structured light pulses to quantum control[16-19]. We have also shown that almost any kind of current patterns can be generated using a combination of a circularly polarized near-infrared (NIR) pulse and a linearly polarized visible light pulse whose phase-front has been structured by a spatial light modulator[17,18]. Each transient current element acts as a subwavelength source of the THz field and a large collection of such transient current elements constitutes an active metasurface that radiates structured, single-cycle THz pulses. Hence, quantum control allows us to generate exotic light modes from a planar semiconductor substrate, bypassing the need for nanostructured materials.

Here, we demonstrate a clear link between the spatial structure of the bichromatic optical fields, the currents they generate, and the spatial structure of the radiated single-cycle THz pulse. Specifically, we use azimuthally-polarized bichromatic optical fields to drive a ring current in gallium arsenide (GaAs). In one part of the experiments, we measure the spatio-vectorial structure of these currents. In the other, we collect the radiated single-cycle THz pulse and map its spatiotemporal properties using electro-optic sampling (EOS). We show that the single-cycle THz pulse radiated from the dynamic ring current has an electric field structure that is azimuthally polarized and that the space- and time-resolved magnetic field has a strong, isolated longitudinal component, which could be of great utility for THz magnetic field spectroscopy. The measured pulse is a "flying doughnut" in THz frequency domain. The link between the incident optical fields, the injected currents, and the radiated 'flying doughnut' THz pulse is illustrated conceptually in Fig. 1a. Here, we note how easily the current structure, and the structure of the radiated single-cycle THz pulse, can be modified to allow the electric and magnetic fields to interchange roles, simply by changing the mode of the incident optical fields.

Having generated a "flying doughnut", we exploit the fact that our THz flying doughnut pulse is propagating through ambient air that contains water vapour. By performing measurements in different humidity levels, we illustrate, for the first time, how absorption features are mapped onto the spatio-temporal structure of a flying doughnut.

Currents were generated in a GaAs substrate under the excitation of two-colour femtosecond laser pulses with characteristics that are described in the Methods section and using an optical set-up sketched in Figure S1. The fundamental pulse has a wavelength of 1480 nm and its second harmonic has a wavelength of 740 nm. They induce two-photon and single-photon transitions in GaAs, respectively. When we use linearly-polarized Gaussian laser pulses, the generated current is uniform and parallel to the local field across the beam profile. Therefore, quantum control driven with two azimuthally polarized vector laser pulses creates a local current vector that is likewise aligned along the local electric fields and, consequently, in a ring structure. In other words, the electric field structure of the driving beams is transferred to the generated current distribution. The spatio-vectorial distribution of the current was measured using a square-apertured (25 µm) low-temperature gallium arsenide (LT-GaAs) detector[16]. The measured current structures are presented in Fig. 1b and clearly resemble a ring current. Again, we note that changing the polarization of the incident optical fields would result in the injection of a radial current arrangement.

We numerically confirm that a dynamic ring current similar to the one presented in Fig. 1b can be a source for a THz flying doughnut by performing finite-difference time-domain (FDTD) simulations. In these simulations, we use a dynamic ring current density as a source and record the spatio-temporal structure of the fields that it radiates. The spatio-temporal structure of the simulated electric field is shown in Fig. 1c. The spatio-spectral structure of the electric fields (Fig. S2) bears a clear resemblance to that expected from a flying doughnut[8,9]. Figure 1d shows the spatio-temporal structure of the simulated magnetic field, which is of sub-cycle duration and whose spatial maximum coincides with the electric field singularity.

Motivated by this prediction, we measure the emitted THz radiation by free-space electro-optic sampling, as depicted in Fig. 2a. For the measurement, the generated THz radiation is collected, collimated and refocused onto a 1-mm-thick zinc-telluride (ZnTe) crystal with a pair of THz lenses (Tsurupica material), as described in Methods. A synchronised second harmonic probe beam (740 nm) is used to sample the THz pulse.

An exemplary THz waveform is shown in Fig. S3a. Figure S3b depicts the power spectral density corresponding to this waveform. We note that the full bandwidth of the THz pulse was not captured in our measurement, as the electro-optic response of the detection crystal (ZnTe) is very weak beyond 5 THz. The detected THz waveform as a function of the relative phase of the two-colour field ($\Delta\varphi_{\omega,2\omega}$) is presented in Fig. 2b. The sinusoidal dependence of the THz peak field on the relative phase ($\Delta\varphi_{\omega,2\omega}$) confirms that the THz pulse arises from quantum control. Therefore, the amplitude and polarity of any structured single-cycle THz pulse generated in this manner can be controlled using the relative phase.

To spatially characterise the THz beam, we raster scan the probe beam with a 200-µm pinhole placed after the ZnTe crystal. The spatiotemporal profile of the THz pulse is obtained by recording the THz waveform at each position in the transverse plane of the beam. Figure 2c shows the measured spatiotemporal structure of the THz pulse emitted from the currents excited by linearly-polarized Gaussian two-colour laser pulses. Clearly, the measured THz beam also has a Gaussian spatial profile and exhibits a constant phase along the transverse direction, as expected from a unidirectional current source.

We then introduce q-plates[20] (orientation depicted in Figure S1(b)) to the experiment to produce two azimuthally- (or radially-) polarized vector pulses for quantum control of ring (or radial) currents in GaAs. The spatiotemporal profile of the THz pulse radiated from a ring current is measured and presented in Fig. 2d. We observe that the phase of the measured THz electric field is flipped along the radial direction and the field almost vanishes at the centre. This corresponds to a doughnut-shaped structure as expected from the calculation in Fig. 1c. Next, we change the polarization of the ω and 2ω light pulses from azimuthal to radial, which generates a radial current. In Fig. 2e, we present the measured space-time map of the radiated THz electric field from the radial current. The result shows a similar phase singularity as seen with azimuthal polarization. It is important to note that we can only measure the transverse components of the radially polarized mode as we use <110>-cut ZnTe crystal. For each of the spatiotemporal measurements, we observe that the electric field profiles are slightly asymmetric. The asymmetry is possibly due to inhomogeneous current generation in GaAs.

We also measure the spatial profiles $E_x$ (*x*, *y*) and $E_y$ (*x*, *y*) of the flying doughnut THz pulses (Fig. 3a and Fig. 3b) radiated from the transient ring current. Spatial maps are measured at the peak of the THz waveform. The far-field spatial profiles closely resemble the two components of an azimuthally-polarized beam. Similar measurements are also performed with radially-polarized light pulses, which are presented in Supplementary Information (Fig. S5). It shows that the positions and polarization of the radiated electric fields are interchanged with respect to those with azimuthal polarization. We see that, by simply changing the linear polarization direction of the Gaussian beam with which we start, we can interchange the roles of the electric and magnetic fields while still retaining the doughnut shape of the radiated pulse. Achieving such flexibility with metasurfaces is not trivial.

The measured spatiotemporal profile of the THz electric field enables us to calculate the space-time structure of the corresponding magnetic field using FDTD simulations. Performing these simulations in cylindrical coordinates means that we only require data from the positive or negative radial coordinates, i.e., half of the scan. The measured electric field structure is presented in Fig. 3c. The strength of the peak electric field is estimated to be 4 kV/cm. The calculated longitudinal magnetic field is presented in Fig. 3d, where the peak magnetic field strength is approximately 0.5 mT. Clearly, the position of the longitudinal magnetic field maximum coincides with the electric field singularity, making it suitable for sensitive control of spin and magnetic systems.

As a first application, we apply the THz flying doughnut to THz time-domain spectroscopy (THz-TDS). Beyond being a very sensitive spectroscopic tool, THz-TDS enables us to obtain direct insight into how an absorption feature might influence the spatio-temporal structure of the emerging pulse, which may not be trivial. Here, we imprint a spectroscopic signature of water vapour onto the generated flying doughnut THz pulse by controlling the humidity level of the air through which it propagates. We consider two situations: ambient air with a relative humidity (RH) of 41% (Fig. 4) and dry air with a relative humidity of 12% (Fig. S6). For the case of ambient air, we clearly observe oscillations after the main peak in the spatiotemporal structure of the measured

THz electric field (Fig. 4a) due to absorption lines of water vapour (Fig. 4b). Figure 4c presents a calculated magnetic field map, where similar oscillatory features are clearly evident. Such absorption lines are also observed in the calculated magnetic field spectral profile (Fig. 4d). We envisage that spatiotemporal THz modes such as the ones we have demonstrated will be highly beneficial for spatially-resolved THz spectroscopy and spectroscopy of toroidal systems[21-23].

In conclusion, we have demonstrated a new approach to metaoptics that eliminates the need for a structured material; instead, we generate transient current structures. Following this approach, we will be able to use tailored two-colour laser fields to excite a single-cycle THz pulse with any mode which is consistent with the material response and with Maxwell's equations. Similar active metastructures can also be generated and controlled in a gas medium[24], a multiphoton extension of what has been demonstrated here in a semiconductor. In fact, high-intensity infrared structured light beams can be applied to gases to drive large current densities and yield high power THz radiation[24-26].

Using this approach, we have created a long sought after, far-field, spatiotemporal mode of light often called a "flying doughnut" in the THz domain. A substantial longitudinal magnetic field transient is present in the generated light mode. Such spatially isolated magnetic impulses are easily synchronised with ultrafast optical excitation and characterisation schemes, and they are important for diagnosing the dynamic response of matter[27-29]. We should mention that these magnetic fields are too weak, as yet, to compete with an accelerator[30].

However, we should distinguish between near and far fields. Ours is a far-field measurement in which the collection efficiency of the optical system has not been fully optimized. Taking the collection and imaging efficiency of our optical system into account, we estimate that the peak magnetic field generated in GaAs would be approximately 10 mT, or roughly 20 times greater than the measured value of 0.5 mT. We note that radiated field can be increased further by irradiating a larger area of the GaAs. Additionally, using a nanostructured material, the THz magnetic field can be enhanced significantly (Supplementary Note 7). It is important to mention here that much larger magnetic fields are available in the near field, as presented in gas breakdown simulations[24]. In the GaAs experiment presented here, the emission per unit area is limited by the need for the THz radiation created within the GaAs to escape through the over-dense plasma. This limit can be overcome by the much more energetic electrons created in gases[24] or in multilayer graphene[31,32].

We have exploited the space-time coupled structure of the flying doughnut for a spectroscopic measurement of the distributed water vapour in ambient air. In this demonstration experiment, we observe that the absorption signature is mapped onto the space-time structure of the measured electric field and calculated magnetic field profiles. Here, we don't observe any change in the spatial structure of the doughnut pulse due to the absorption. We expect that for a nonlinear interaction, the interaction process will be diagnosed by measuring the change in spatial structure of the doughnut pulse. This is a characteristic of flying doughnut pulses and it can be used for high-temporal-resolution spectroscopy of materials in response to either electric or magnetic fields.

Finally, while single-cycle light pulses have been available at THz frequencies for decades, it is only recently that technology has advanced sufficiently to produce single-cycle pulses in the visible and near infrared. This laser technology can almost certainly be extended to the mid-infrared (MIR)[33] where the $CO_2$ is an important gain medium. $CO_2$ lasers are polarization insensitive. Therefore, an azimuthally polarized pulse can be generated at low energy and then amplified to Joule-level picosecond pulses[34-36]. Hollow-core fibre technology can already produce near single-cycle azimuthally or radially polarized pulses at 800 nm and this can almost certainly be scaled to Joule-level pulses in the MIR[34-38]. Thus, we expect that $CO_2$ laser technology will give rise to high-power 'flying doughnut' pulses with frequencies of ~30 THz and isolated magnetic fields that could approach amplitudes of $10^5$ Tesla with a wavelength-scale focal spot. An isolated $10^5$ Tesla and greater fields are only available from astronomical objects such as neutron stars ($10^7 - 10^9$ Tesla) or magnetic white dwarfs ($10^2 - 10^5$ Tesla). $10^5$ Tesla fields are important because $10^5$ Tesla is approximately the field at which the Atomic Unit of Energy (the energy that characterizes atomic hydrogen) equals the energy of an electron undergoing cyclotron motion i.e., $\hbar\omega=\hbar qB/m$, where $q$ and $m$ are the electron charge and mass while $B$ is the magnetic field strength[39]. Even spectroscopy involving the ground state of the hydrogen atom is severely modified. For energies of this scale, 30 THz is essentially a static field.

## Methods

**Experimental Setup:** The schematic of the full experimental setup is shown in Fig. S1 in Supplementary Information. The main fundamental beam centered at 1480 nm is obtained from a commercial optical parametric amplifier (TOPAS, Light Conversion) pumped by amplified 5mJ, 800 nm femtosecond pulses (Legend system, Coherent) operating at 1 kHz repetition rate. The fundamental beam is spatially filtered by passing it through a hollow core fiber kept in ambient air. The spatially filtered pulse is then transmitted through a beta-barium-borate (BBO) crystal for second harmonic (740 nm) generation. Subsequently, a dichroic mirror is used to separate the two pulses. The two pulses are independently manipulated in each arm of the interferometer. Their energy is controlled by a combination of a half-wave plate and a polarizer. A liquid crystal wave plate (q-plate) is used to generate azimuthal polarization from the linearly polarized beam. After combining the two beams with a dichroic mirror, they propagate collinearly. Finally, a lens is used to focus the two pulses onto the gallium arsenide (GaAs) substrate to excite transient ring currents. The intensities of the $\omega$ and $2\omega$ beams were kept at approximately $2 \times 10^{11}$ W/cm$^2$ and $4 \times 10^9$ W/cm$^2$, respectively. A flip mirror mount is used to direct the bichromatic beam to either a current measurement or THz generation. A small part of the $2\omega$ beam is used as the probe pulse for EOS of the terahertz waveforms. Using a pellicle beam splitter, the unfocused probe beam was aligned collinearly with the THz beam.

**Simulations:** The EOS measurements provide the space-time structure of a single slice of the THz pulse. We perform finite-difference time-domain (FDTD) simulations to obtain the space-time structure of the complementary magnetic fields, along with an estimation of their amplitude. The FDTD code is home-built and is implemented in cylindrical coordinates. We assume that the fields are independent of the azimuthal coordinate, which reduces the simulation space to two dimensions. We use a uniform grid with $\Delta r = \Delta z = 5$ μm and $\Delta t = 11.7$ fs. The raw measured electric fields are used as a source in the simulations, and the three field components ($E_\varphi, B_r, B_z$) are recorded after the pulse has propagated 4 mm.

## Acknowledgements

The authors would like to thank Prof. Ebrahim Karimi and his group for providing with the q-plates. The authors are also grateful for important discussions with Saroj Tripathi, Andre Staudte, Chunmei Zhang, David Purschke. This research was supported by the Natural Sciences and Engineering Research Council of Canada (NSERC) Discovery Grant Program, the Canada Research Chairs program, the United States Defense Advanced Research Projects Agency ("Topological Excitations in Electronics (TEE)", agreement #D18AC00011), and the United States Army Research Office (award #: W911NF-19-1-0211).

## Author contributions

P.B.C. conceived the idea and supervised the experiments. K.J. designed the setup, performed the measurements, analysed the data and wrote the first draft of the manuscript. S.S. performed the calculations. All authors discussed the results and contributed to the manuscript.

## Competing interests

The authors declare no competing financial interests.

## Data availability

The data that support the findings of this study are available from the corresponding author upon reasonable request.

## Additional information

**Supplementary information** is available for this paper. Correspondence and requests for materials should be addressed to K.J.

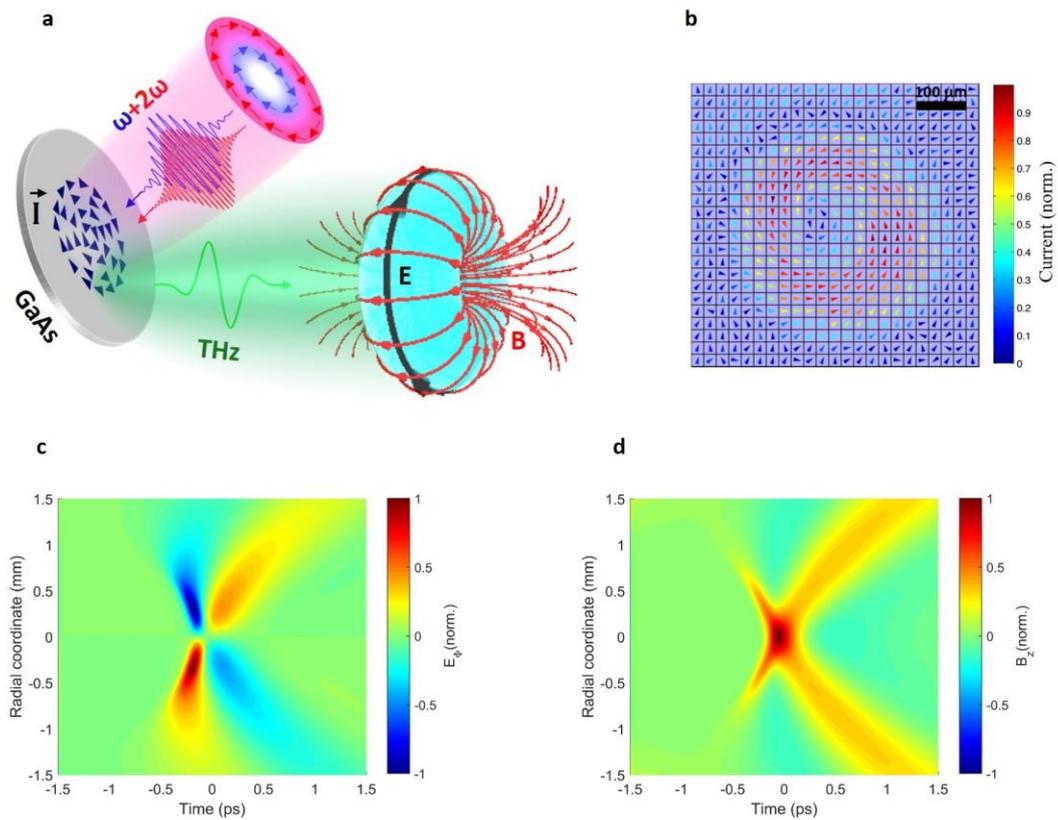

**Fig. 1 | Dynamic ring current radiates 'Flying doughnut' terahertz pulse. a,** Illustration of 'Flying Doughnut' pulse generation. Two azimuthally-polarized vector pulses generate transient ring currents in GaAs. The radiation from the rapidly oscillating ring currents is a single-cycle THz pulse with toroidal topology, a 'Flying Doughnut'. **b,** Spatio-vectorial distribution of ring current measured with a single pixel current detector. Combining the detected x and y components (Fig. S2) of the current results in spatial mapping of the ring current. **c,** Spatio-temporal structure of the radiated electric field ($E_\varphi$) simulated using a dynamic ring current density source. **d,** Simulated magnetic field ($B_z$) map of the emitted THz pulse.

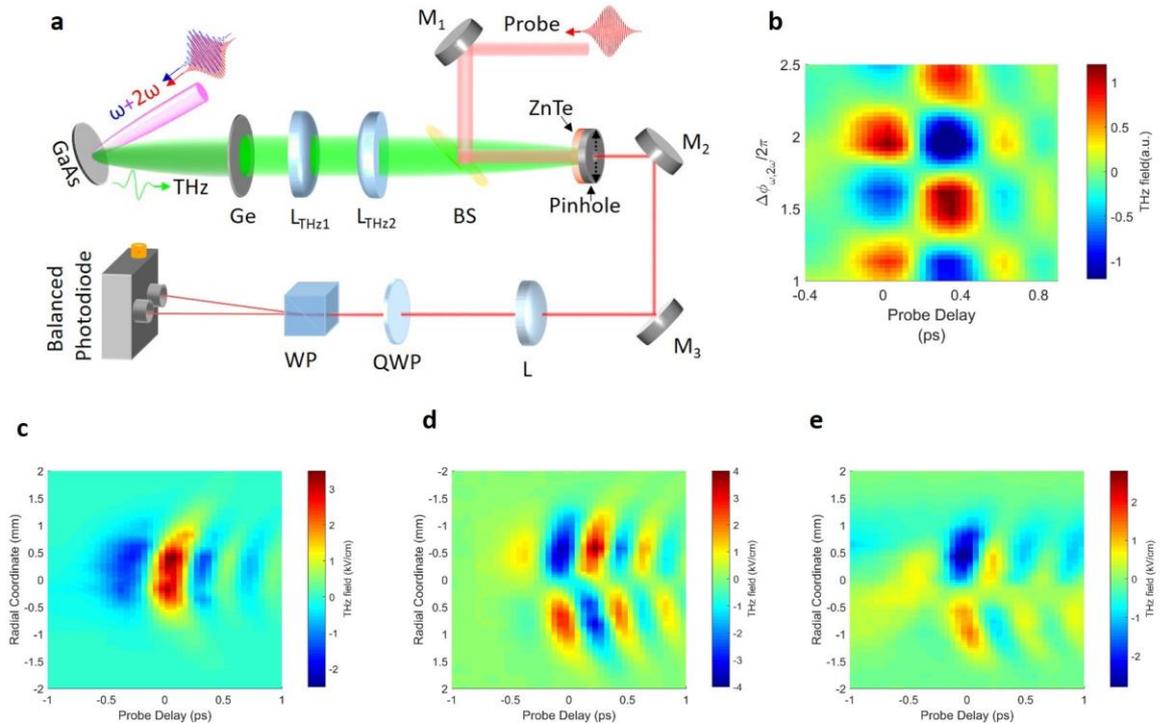

**Fig. 2 | Terahertz emission from two-colour injected currents in GaAs. a,** Schematic of the experimental setup for measuring terahertz radiation from transient currents excited in GaAs by synthesised two-colour fields. GaAs: gallium arsenide

substrate, Ge: germanium wafer, $M_1$-$M_3$: metallic mirrors, L: focusing lens, $L_{THz1}$-$L_{THz2}$: THz lenses, QWP: quarter-wave plate, WP: Wollaston prism, BS: pellicle beam splitter, ZnTe: zinc telluride crystal. **b**, THz field as a function of two-colour phase ($\Delta\varphi_{\omega,2\omega}$) and probe time delay. Terahertz waveforms are recorded at different values two-colour phase. Spatiotemporal maps of the measured THz electric field when **c,** linearly polarized **d,** azimuthally-polarized and **e,** radially-polarized two-colour pulses are applied.

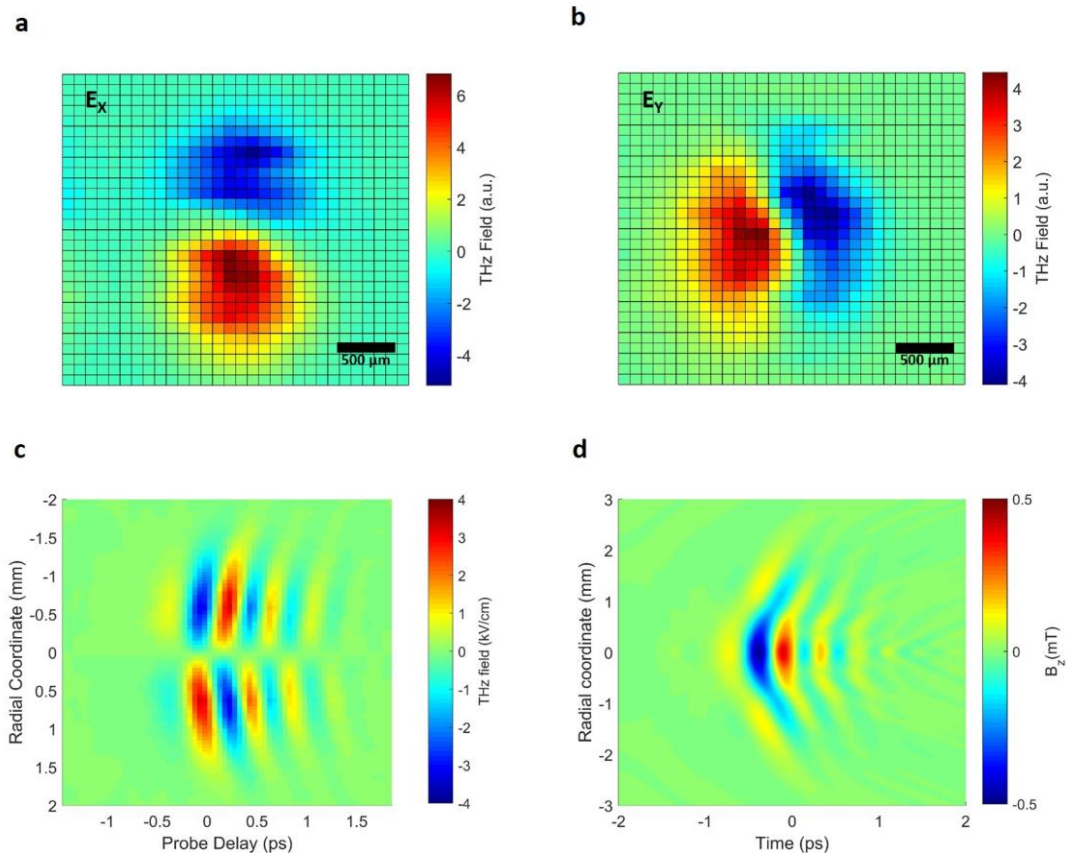

**Fig. 3 | Measurements of 'Flying doughnut' terahertz pulses.** Far-field spatial maps of the flying doughnut THz pulse **a,** $E_x$ ($x$, $y$) and **b,** $E_y$ ($x$, $y$), radiated from the ring current. Spatial profiles are measured at the peak of the THz waveform. **c,** Spatiotemporal structure of the electric field of the flying doughnut THz pulse. Measured data is symmetrized for magnetic field calculation. **d,** Space-time structure of the **c**alculated longitudinal magnetic field ($B_z$) of the flying doughnut pulse. The magnetic field is calculated from the measured electric field data using Maxwell's equations.

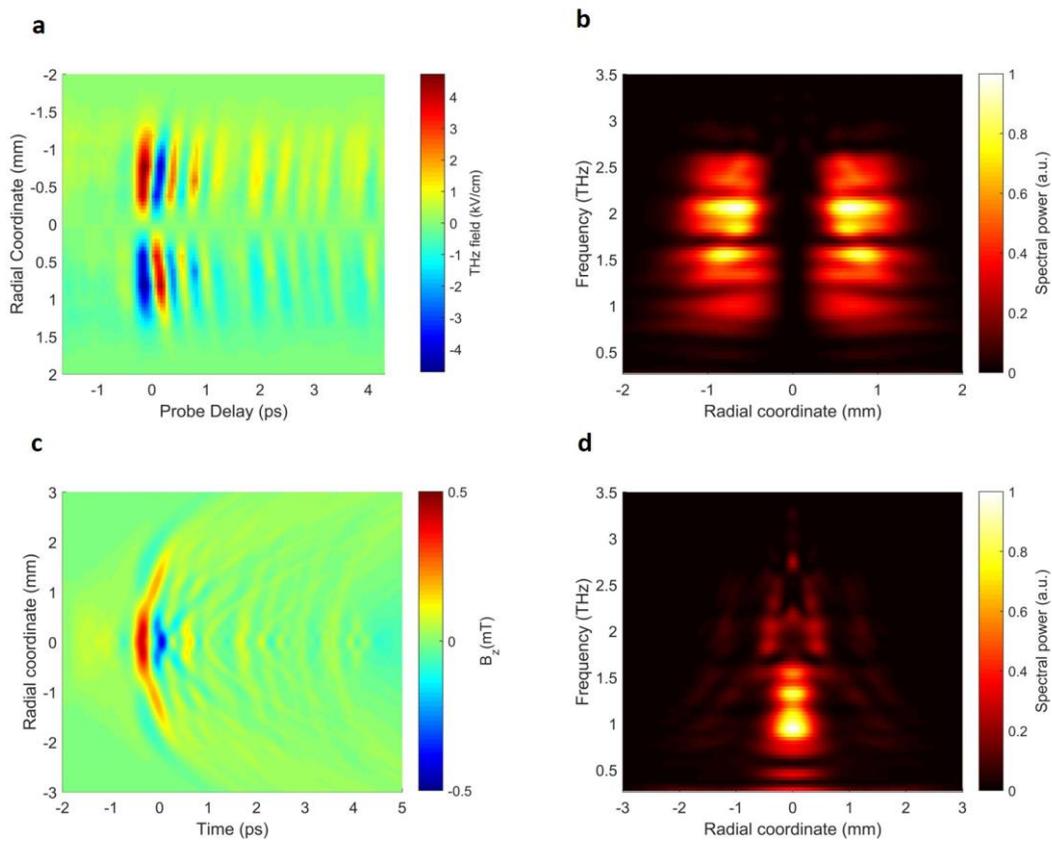

**Fig. 4 | Spectroscopy using flying doughnut pulse. a,** Measured electric field of flying doughnut pulses in humid air (RH 41%) condition. **b,** Space-frequency representation of the measured electric field. **c,** Calculated magnetic field ($B_z$) structure of the flying doughnut pulse in humid air. **d,** Spatiospectral map of the magnetic field $B_z$.

# Supplementary Information

## Flying doughnut terahertz pulses generated from semiconductor currents

Kamalesh Jana[1*], Yonghao Mi[1], Søren H. Møller[1], Dong Hyuk Ko[1], Shima Gholam-Mirzaei[1], Daryoush Abdollahpour[1,2], Shawn Sederberg[3], Paul B. Corkum[1*]

[1]*Joint Attosecond Science Laboratory, University of Ottawa and National Research Council Canada, 25 Templeton Street, Ottawa, ON K1N 6N5 Canada*

[2]*Department of Physics, Institute for Advanced Studies in Basic Sciences (IASBS), Zanjan 45137-66731, Iran*

[3]*School of Engineering Science, Simon Fraser University, 8888 University Drive, Burnaby, BC, V5A 1S6, Canada*

*e-mail: *kjana@uottawa.ca* (K.J.); *pcorkum@uottawa.ca* (P.B.C.)

## Supplementary note 1 – Experimental setup for current and THz measurements

The full experimental setup is schematically depicted in Fig. S1(a). The two-colour ($\omega$ and $2\omega$) pulses are passed through an interferometer that allows the relative phase to be adjusted by a piezo controller.

For the current measurements, a flip mirror mount is positioned to reflect the combined $\omega$ and $2\omega$ pulses. A focusing lens ($f$ = 200 mm) is used to focus the combined pulses onto the LT-GaAs optoelectronic detector. The single-pixel detector is raster scanned to measure the spatial distribution of the currents at the image plane.

For THz measurements, flip mirror mount is set such that the two-colour pulses do not reflect from the mirror. The combined pulses are now gently focused (spot diameter 700 µm) onto the 110-µm-thick GaAs substrate using a focusing lens to generate currents. The intensities of the $\omega$ and $2\omega$ beams on the sample were $2 \times 10^{11}$ W/cm$^2$ and $4 \times 10^{9}$ W/cm$^2$, respectively. THz pulses are radiated in both the backward and forward directions from the transient ring current. However, the THz pulse propagating through the thick GaAs substrate will be significantly suppressed by the free-carrier plasma generated by the $\omega$ beam. Therefore, we collect the THz pulse radiated from the front side into air (reflection geometry) by a THz lens ($f$ = 75 mm) and subsequently refocus it onto the electro-optic crystal (<110>-cut ZnTe) by another THz lens ($f$ = 100 mm). In front of the first THz lens, we place a 0.5-mm-thick uncoated germanium substrate to block the residual $\omega$ and $2\omega$ pump pulses and filter out only THz pulses. We pinhole with the ZnTe crystal to spatially resolve the THz radiation. The assembly (ZnTe and pinhole) was placed on a two-dimensional ($x$-$y$) linear stage for raster scanning the probe. By doing this we rule out the additional effect in the measurements due to the non-uniformity of the ZnTe crystal. In order to mapout the far field spatial profiles of peak THz fields $E_x$ ($x$, $y$) and $E_y$ ($x$, $y$) we used a THz polarizer to select the polarization and raster scanned the pinhole across the probe. We also rotated the orientation of ZnTe crystal accordingly such that it measures $E_x$ and $E_y$ efficiently. THz is partly absorbed by the distributed water vapor present in the ambient air [Relative humidity (RH) 41%]. The THz part of the setup is enclosed and purged with dry air (achieved RH 12%) to reduce the water vapor absorption of THz pulse.

Peak THz field was calculated from the intensity modulation in balance detected EOS signal [1] i.e., $\sin^{-1}\left(\frac{\Delta I}{I}\right) = \frac{2\pi n_0^3 r_{41} t_{ZnTe} E_{THz} L}{\lambda_0}$ , where $L$ is ZnTe thickness, $r_{41}$ is EO coefficient of ZnTe, $n_0$ is the refractive index of ZnTe at probe wavelength, $\lambda_0$ is probe wavelength, $t_{ZnTe}$ is coefficient of transmission. We also estimated peak THz electric field from the expression $E_{THz} = \sqrt{\frac{2W}{c\varepsilon_0 A\tau}}$, where $W$ is the THz pulse energy, $c$ is the speed of light in vacuum, and $\varepsilon_0$ is the vacuum permittivity. While all the parameters are known here, the THz pulse energy is assumed to be 40 pJ considering $10^{-6}$ conversion efficiency [2]. The estimated values of the THz peak field using these two methods really match well.

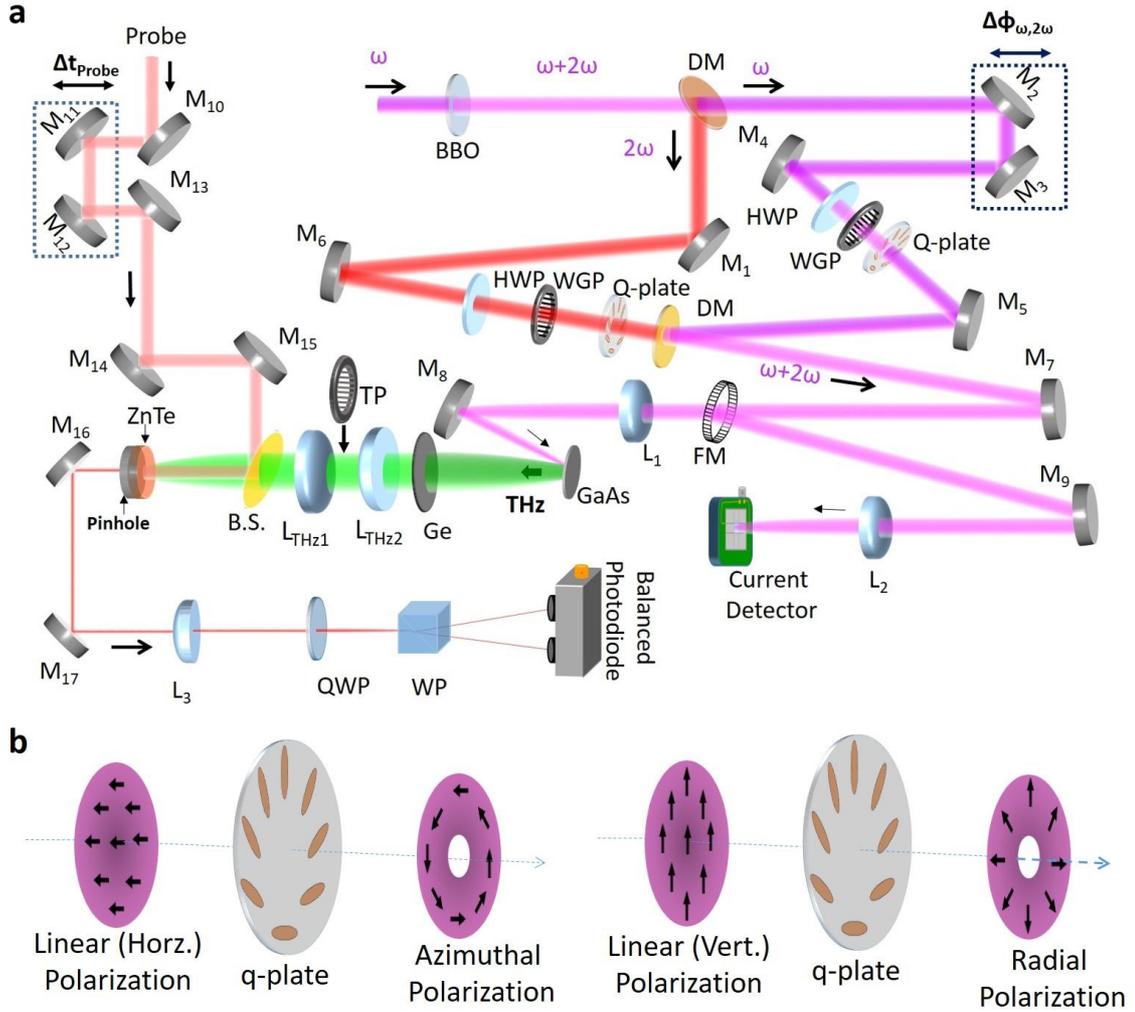

**Fig. S1 | Schematic of the experimental setup for THz and current measurements.**
(a) Spatially filtered $\omega$ beam is passed through the BBO two generate $2\omega$ pulse. Two pulses then travel through the two different arms of a two-colour interferometer and are finally combined by a DM. BBO: beta-barium borate, DM: dichroic mirror, $M_1$-$M_{17}$: metallic mirrors, FM: flipper mirror, $L_1$-$L_3$: focussing lenses, $L_{THz1}$-$L_{THz2}$: THz lenses, TP: THz polarizer, HWP: half-wave plate, WGP: wire-grid polarizer, QWP: quarter-wave plate, WP: Wollaston prism, Q-plate: q-wave plate, GaAs: gallium arsenide substrate, Ge: germanium wafer, BS: pellicle beam splitter, ZnTe: Zinc telluride crystal. (b) Generation of azimuthal and radial polarizations from linear polarization using q-plate.

## Supplementary note 2 – X and Y-components of the measured ring current

Our LT-GaAs current detector is direction sensitive. We measure the $x$ and $y$-components of the current separately by rotating the detector accordingly. Combining the measured $I_x$ and $\underline{I}_y$ enables spatial mapping of the ring current (presented in Fig. 1b).

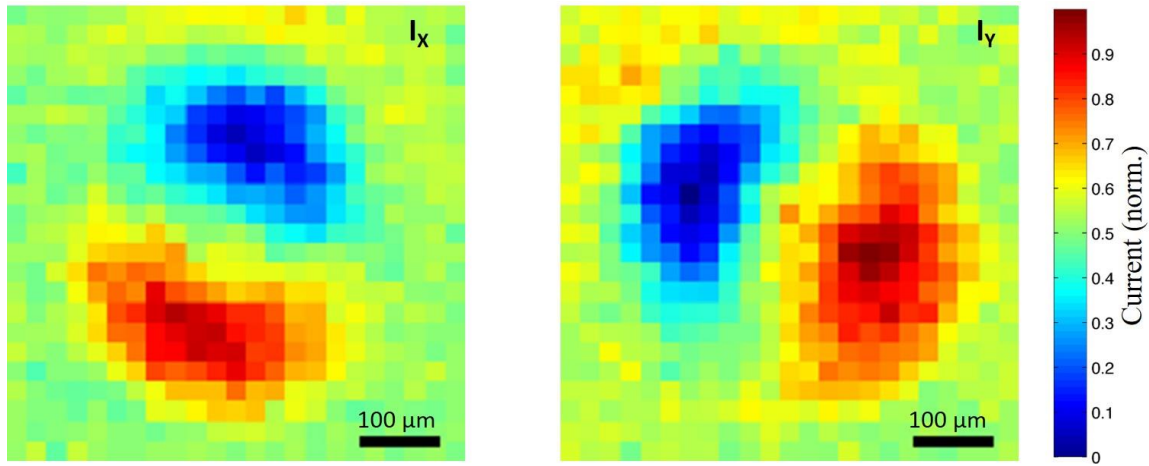

**Fig. S2 | Ring current measurement.** X (left) and Y (right) components of the ring current vector are measured using single-pixel LT-GaAs based optoelectronic detector.

## Supplementary note 3 – Spatiospectral structure of the calculated flying doughnut pulse

We simulated the space-time structure of the radiated electric field (Fig. 1c) and magnetic field (Fig. 1d) from the ring current. The frequency domain representations of the calculated electric and magnetic fields are depicted in Fig. S3.

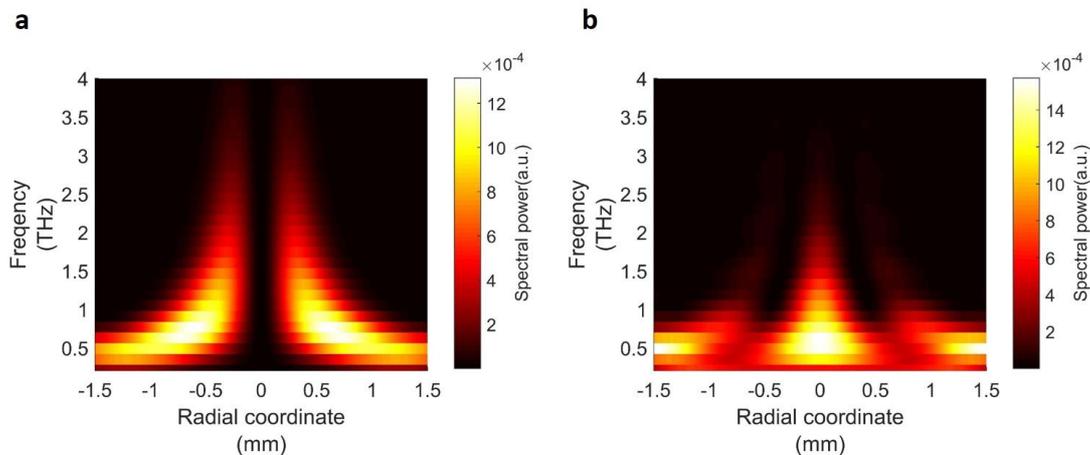

**Fig. S3 | Spatiospectral maps,** Space-frequency representation of the simulated (a) electric field and (b) magnetic field of the flying doughnut pulse from the ring current.

## Supplementary note 4 –THz waveform radiated from GaAs under two-colour excitation

Rapidly oscillating two-colour injected current in GaAs radiates terahertz electromagnetic radiation. A typical THz waveform is presented in Fig. S4.

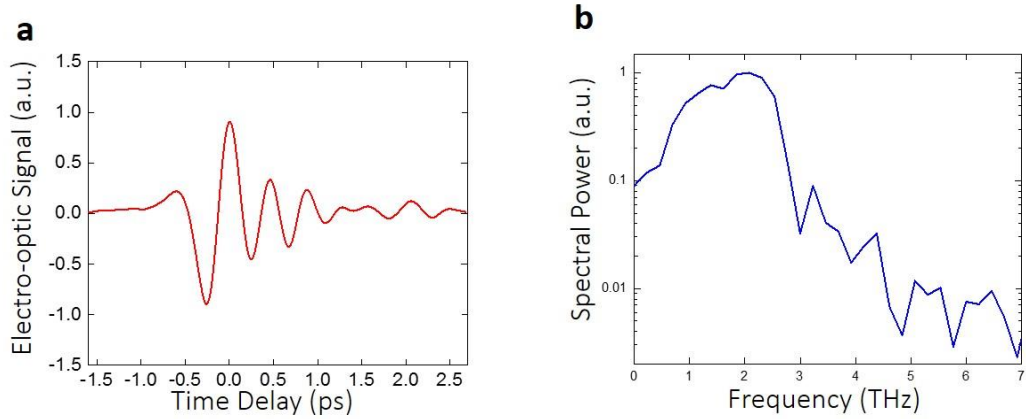

**Fig. S4 | THz pulse from GaAs under two-colour excitation.** (a) Time-domain and (b) frequency domain representation of a typical THz waveform.

## Supplementary note 5 – Far-field profiles of the radiated THz pulse from radial current

We measured spatial profiles $E_x(x, y)$ and $E_y(x, y)$ of the radiated electric field from the radial current. The spatial ($x$-$y$) maps of the electric fields were measured at the peak of the radiated THz pulse. As we used <110>-faced ZnTe crystal we can only measure transverse profiles of the radial THz mode.

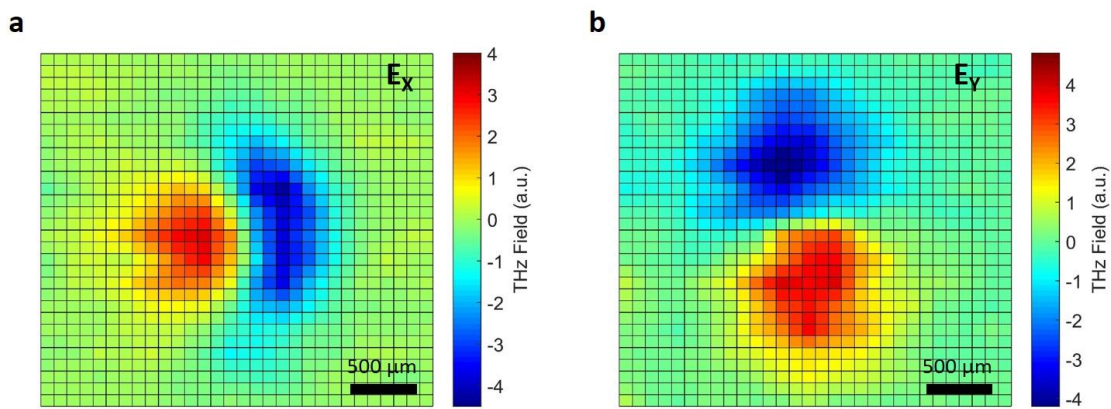

**Fig. S5 | Spatial profiles of THz pulse radiated from the radial current.** Spatial profiles of the two transverse components, (a) $E_x(x, y)$ and (b) $E_y(x, y)$ of the THz pulse radiated from the radial current distribution.

## Supplementary note 6 – Spectroscopy with flying doughnut pulse in dry air condition

We measured flying doughnut pulse in dry air condition (RH 12%). The electric and magnetic field data is presented in S6

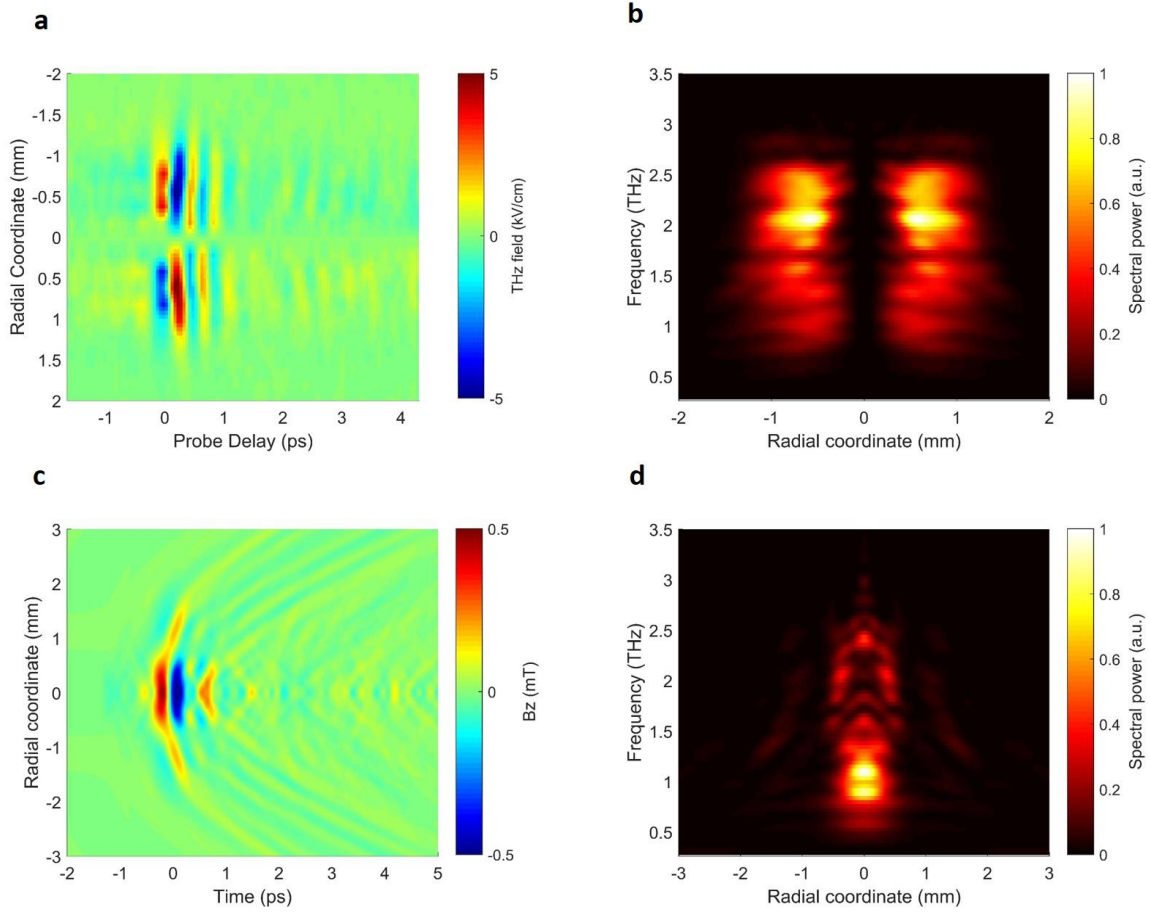

**Fig. S6 | Measurements of flying doughnut pulse in dry air condition.** (a) Measured electric field of flying doughnut pulses in dry air (RH 12%) condition. (b) Space-frequency representation of the measured electric field. (c) Calculated magnetic field ($B_z$) structure of the flying doughnut pulse in dry air. (d) Spatiospectral map of the magnetic field $B_z$.

## Supplementary note 7 – Optimization and direct measurement of THz magnetic field

In this section, we discuss how the radiated THz magnetic field can be increased. First, improving the collection and imaging efficiency of our optical system would considerably enhance the measured THz field. We use two THz lenses to collect and refocus the THz radiation. Having a first lens of F#3, does not collect the radiated light efficiently while with the second lens (F#4) we measure focused radiation with a focal spot diameter of 2.5 mm, thereby magnifying the 700 μm source by M=3.5. Finally, we filter out THz beam with 0.5 mm thick germanium substrate. The THz transmission of the germanium is ∼ 40%. If the radiation were collected and imaged efficiently we predict that we would have measured ∼10 mTesla for the peak magnetic field.

Second, we have only used 50 µJ of the mJ 1.48 µm light available. By using more laser power and illuminating larger area while keeping the same power density, we will generate a larger THz field. For example, were we to irradiate a full 2.5 mm, we predict a radiated field of ~ 40 mTesla.

Third, using the nano-structured metals the THz magnetic field can be enhanced further. A factor of six enhancements was reported recently [3, 4]. Combining all these factors, we expect that it will be possible to generate a few 100 mT THz magnetic field from quantum-controlled currents excited in GaAs. Further enhancement will require other semiconductors or a more energetic 1.48 µm source. Tesla scale THz magnetic fields seem realistic from GaAs.

We also envisage that a terbium gallium garnet (TGG) crystal could be used in the place of ZnTe for direct measurement of the temporal structure of the THz magnetic field [5]. The total Faraday rotation can be calculated using formula **$\theta = VBL$**; where *V* is the Verdet constant of TGG, *B* is the magnetic field strength and *L* is the effective thickness of the TGG.

For our present experimental setup, 12 mm is the depth of the field (for 1 THz) and we predict $\theta$ =480 micro-radians. Although our current measurement uses electro-optic sampling and measures the maximum rotation 100 milliradians, we have a signal to noise ratio of 250:1. Technically, we could resolve a rotation of 480 micro-radians and we could easily resolve a signal (with enhanced magnetic field) corresponding to 10 milliradians.